\documentclass[aps,pre,nofootinbib,superscriptaddress,nopacs,amsmath,amssymb,final,letterpaper,notitlepage]{revtex4-1}

\usepackage[utf8]{inputenc}
\usepackage{calc}
\usepackage{graphicx}
\usepackage{amsmath,amssymb,amsthm}
\usepackage{bbold}
\usepackage{bm}
\usepackage{color}
\usepackage[dvipsnames]{xcolor}
\usepackage{enumitem}
\usepackage{tikz}
\usepackage{float}
\usepackage[percent]{overpic}
\usepackage{parskip}
\usepackage{adjustbox}
\usepackage{multirow}

\makeatletter
\renewcommand{\section}{%
  \@startsection{section}{2}{0pt}%
  {\baselineskip}% Space above
  {0.5\baselineskip}% Space below
  {\normalfont\normalsize\bfseries\raggedright}% Left-aligned, bold, normal size
}
\makeatother

\makeatletter
\renewcommand{\subsection}{%
  \@startsection{subsection}{2}{0pt}%
  {\baselineskip}% Space above
  {0.5\baselineskip}% Space below
  {\normalfont\normalsize\bfseries\raggedright}% Left-aligned, bold, normal size
}
\makeatother

\makeatletter
\renewcommand{\subsubsection}{%
  \@startsection{subsubsection}{2}{0pt}%
  {\baselineskip}% Space above
  {0.5\baselineskip}% Space below
  {\normalfont\normalsize\bfseries\raggedright}% Left-aligned, bold, normal size
}
\makeatother

\let\oldabstract\abstract
\let\endoldabstract\endabstract
\renewenvironment{abstract}
  {\oldabstract\vspace{-\baselineskip}} % Reduce space before abstract
  {\endoldabstract\vspace{-\baselineskip}} % Reduce space after abstract

\usepackage{enumitem}

\newcommand{\abs}[1]{\vert #1\vert}

\def\multiset#1#2{\ensuremath{\left(\kern-.3em\left(\genfrac{}{}{0pt}{}{#1}{#2}\right)\kern-.3em\right)}}

\usepackage[colorlinks=true,linkcolor=blue,urlcolor=blue,citecolor=blue,anchorcolor=blue]{hyperref}

\begin{document}
\title{Analyzing Students’ Emerging Roles Based on Quantity and Heterogeneity of Individual Contributions in Small Group Online Collaborative Learning Using Bipartite Network Analysis}

\author{Shihui Feng}
\email{shihuife@hku.hk}
\affiliation{Faculty of Education, University of Hong Kong, Pok Fu Lam Road, Hong Kong, China}
\affiliation{Institute of Data Science, University of Hong Kong, Pok Fu Lam Road, Hong Kong, China}

\author{David Gibson}
\affiliation{Curtin University, Perth, Australia}
\affiliation{UNESCO Chair on Data Science in Higher Education Learning \& Teaching}

\author{Dragan Gašević}
\affiliation{Faculty of Information Technology, Monash University, Clayton, VIC, Australia}

\begin{abstract}

Understanding students’ emerging roles in computer-supported collaborative learning (CSCL) is critical for promoting regulated learning processes and supporting learning at both individual and group levels. However, it has been challenging to disentangle individual performance from group-based deliverables. This study introduces new learning analytic methods based on student–subtask bipartite networks to gauge two conceptual dimensions---quantity and heterogeneity of individual contribution to subtasks---for understanding students’ emerging roles in online collaborative learning in small groups. We analyzed these two dimensions and explored the changes of individual emerging roles within seven groups of high school students (N = 21) in two consecutive collaborative learning projects. We found a significant association in the changes between assigned leadership roles and changes in the identified emerging roles between the two projects, echoing the importance of externally facilitated regulation scaffolding in CSCL. We also collected qualitative data through a semi-structured interview to further validate the quantitative analysis results, which revealed that student perceptions of their emerging roles were consistent with the quantitative analysis results. This study contributes new learning analytic methods for collaboration analytics as well as a two-dimensional theoretical framework for understanding students’ emerging roles in small group CSCL.

\end{abstract}

%%%%%%%%%%%%%%%%%%%%%%%%%%%%%%%%%%%%%%%%%%%%%%%%%%%
%%%%%%%%%%%%%%%%%Intro%%%%%%%%%%%%%%%%%%%%%%%%%%%%%
%%%%%%%%%%%%%%%%%%%%%%%%%%%%%%%%%%%%%%%%%%%%%%%%%%%

\maketitle

\section{Introduction}

Collaborative learning enables a group of students to learn through an interactive and co-constructive process (Dillenbourg, 1999; Laal \& Laal, 2012). With the support of information and communication technologies, computer-supported collaborative learning (CSCL) has various forms, spanning from small group collaboration consisting of three to four students to large-scale, wiki-based open collaboration. Large-scale CSCL through online discussion forums or wiki-based platforms is effective in supporting unstructured, decentralized, and temporal collaborative learning activities, as centralized coordination in a large group tends to be costly and challenging. In contrast, small group online collaborative learning provides a setting in which a few students work closely with each other towards a common objective over a certain period. Students’ participative stances and collaboration competency can differ between large and small group collaborations (Strijbos \& De Laat, 2010). \\

There are two types of roles involved in small group CSCL: 1) emerging and 2) scripted (Strijbos \& Weinberger, 2010; Kollar et al., 2006). Emerging roles refer to those developed dynamically and spontaneously through the learning processes, while scripted roles refer to those assigned or expected by instructors prior to collaboration. In a group collaboration, students’ emerging roles can be reflected by their individual contributions in the learning process (Strijbos \& Weinberger, 2010). Neglecting to understand and monitor students’ emerging roles can potentially lead to team-level issues including social loafing and a lack of team development (Hansen, 2006; Worsley et al., 2021). Process regulation and effective teacher feedback rely on a good understanding of students’ emerging roles in learning processes (Coll et al., 2014; Järvelä \& Järvenoja, 2011; Malmberg et al., 2022). At the individual level, it is critical for students to contribute to the collaborative learning process by actively performing the related learning activities and interacting with other members in order to activate their own cognitive systems and gain new knowledge (Dillenbourg, 1999; Nokes-Malach et al., 2015). \\

Previous studies on students’ emerging roles have been mainly based on student social interaction data in large online collaborations (e.g., online forums; Saqr \& López-Pernas, 2022; Kim \& Ketenci, 2019; Marcos-García et al., 2015). Several studies used student communication log data to study their emerging roles in small groups in CSCL (Strijbos \& De Laat, 2010; Dowell et al., 2019; Dowell et al., 2020). Qualitative analysis of content data is costly and impractical to adopt at scale, and the heuristic text mining approach used for analyzing communication log data can suffer from accuracy and interpretability issues (Talib et al., 2016).
This study introduces a bipartite network analysis approach to study students’ emerging roles in small group collaboration learning, based on the quantity and heterogeneity of their engagement with subtasks — the artefacts predefined by teachers to scaffold the collaborative learning processes. It is important to analyze and combine both the quantity and heterogeneity of subtasks worked on by a student in a CSCL process (Weinberger \& Fischer, 2006). Quantity of individual contribution, in terms of the total number of interactions on subtasks, is a direct but partial reflection of how much a student contributes to a group task. This partial measure might entail several contributions to only one task or several interactions across all possible tasks, a variation requiring the second dimension of heterogeneity. Heterogeneity of contribution, on the other hand, reflects the extent to which a student engages in a variety of subtasks involved in a project. Heterogeneity of individual contribution is an important sub-measure to reflect how thoroughly a student has engaged with different cognitive aspects involved in a project; however, it is understudied.
Understanding students’ emerging roles based on the task quantity and task heterogeneity of individual contributions can help instructors and learners in three ways:
\begin{enumerate}
    \item It can assist in addressing the “free riding” problem of attribution — who did what on the team; for example, identifying students who made significant contributions and those who took a “free ride” (Janssen et al., 2007).
    \item It can provide a foundation for instructional redesign; for example, strengthening the features needed to develop a socially shared collaborative learning environment that encourages both depth and breadth of learning by all participants (Hadwin \& Oshige, 2011; Järvelä \& Hadwin, 2013).
    \item It can assist in understanding and quantifying the depth and breadth of individual learning in small group CSCL.
\end{enumerate}

In this study, we used bipartite network analysis to analyze the quantity and heterogeneity of individual contributions based on the student–subtask interactions in small group collaborative learning processes. The automated bipartite network analysis approach proposed here makes machine scoring and analysis feasible, with the benefits of providing a basis for timely feedback to teams during and at the conclusion of a team project. Bipartite network analysis models the relationships between two different sets of nodes (Newman, 2018). This differs from traditional one-mode networks that focus on relationships between entities of the same type (e.g., student–student interactions). Modelling social interactions among three or four students in small group online collaborative learning requires new network-based approaches, as the network of each group tends to be a fully connected graph with a simple topology (e.g., everyone interacts with everyone else at least once). Bipartite network analysis is particularly valuable in these situations for providing evidence for small group collaboration analytics. Bipartite network analysis can leverage the engagement data of students with subtasks to capture student engagement patterns in the learning processes and provide an opportunity to derive data-based evidence of students’ emerging roles. The bipartite network analysis methods used in this study can be adapted for investigating students’ emerging roles in a variety of collaborative learning contexts. By addressing the three questions below, this study aims to advance collaboration analytics methodology and contribute to the theoretical understanding of emerging roles in CSCL.
\begin{itemize}[leftmargin=0.7em]
    \item[]\textbf{RQ1:} How can we identify students’ emerging roles based on the quantity and heterogeneity of contributions in CSCL processes using bipartite network analysis?
    \item[]\textbf{RQ2:} How does the assigned leadership role interact with the emerging roles informed by the quantity and heterogeneity of individual contributions?
    \item[]\textbf{RQ3:} To what extent is the qualitative evidence from student interview responses about their emerging roles through the CSCL experience consistent with the quantitative findings?
\end{itemize}

\section{Literature Review}
\subsection{Emerging Roles in CSCL}
The Zone of Proximal Development (ZPD) theory supports the idea that individuals will not be able to achieve certain goals without support from others (Vygotsky, 1978) and is considered one of the theoretical foundations for understanding the benefits of collaborative learning. Dillenbourg (1999) argued that collaborative learning stimulates learning mechanisms through certain forms of interactions among group members, instead of being a learning mechanism or method by itself. There is no guarantee that learning at the individual level will occur in CSCL, regardless of the completion of the group tasks (Dillenbourg, 1999). Successful collaborative learning not only requires intersubjective meaning-making and co-construction of knowledge at the group level but also requires active contribution from each member at the individual level (Järvelä \& Järvenoja, 2011; Peterson \& Roseth, 2016). Based on the constructivist theory (Piaget, 1976), the occurrence of learning at the individual level requires individuals to actively engage in the process and contribute to the group collaboration. Understanding students’ emerging roles based on their individual contributions is thus considered critical in CSCL. This further raises a key question of how to analyze students’ emerging roles based on their individual contributions using trace data in collaborative learning processes.\\

Compared to scripted roles with assigned tasks and duties known prior to beginning a CSCL project, emerging roles are developed spontaneously during the self-organizing collaborative learning processes (Strijbos \& Weinberger, 2010). Based on the conceptual framework proposed by Strijbos and De Laat (2010), students’ emerging roles can be seen at three levels, including the micro-level roles that focus on process- or product-oriented tasks, the meso-level roles that reflect the patterns of participatory behaviour, and the macro-level roles that relate to the stances of individuals towards collaborative learning. Within the three levels, product-oriented roles are related to the accomplishment of specific tasks, while the process-oriented roles are related to the task management activities for facilitating the collaboration processes (Strijbos \& De Laat, 2010).\\

Previous studies have mostly focused on analyzing students’ process-oriented emerging roles by analyzing their social interaction patterns in large-group collaborative learning (e.g., online discussions) using network analysis, constructing learner profiles based on structural properties and classifying them into different categories of emerging roles using predefined thresholds or data-driven clustering methods (Aviv et al., 2003; Saqr \& López-Pernas, 2022; Kim \& Ketenci, 2019; Saqr et al., 2018; Marcos-García et al., 2015; Turkkila \& Lommi, 2020; Medina et al., 2016). Several studies have analyzed students’ emerging roles in small group CSCL based on the communication content generated by students in the learning processes using various methods, including qualitative coding, computational linguistic techniques, text mining methods, and epistemic network analysis (Dowell et al., 2019, 2020; Ferreira et al., 2022; Xie et al., 2018; Swiecki, 2021; Swiecki \& Shaffer, 2020). Strijbos and De Laat (2010) proposed a conceptual framework describing eight types of participative stances along the dimensions of group size, orientation, and effort. Within small group collaborations, they proposed four types of roles — over-rider, ghost, free rider, and captain, along the dimension of orientation and efforts — with a qualitative coding approach to analyze student transcriptions of their group communications (Strijbos \& De Laat, 2010). However, the qualitative coding involved in analyzing student communication content is costly and challenging for large-scale datasets. Additionally, the heuristic text mining approaches used in previous studies (e.g., LDA topic modelling) require subjective decisions, which can lead to difficulty gauging and evaluating the accuracy and interpretability of the analytical results (Talib et al., 2016).

\subsection{Bipartite Network Analysis}
Networks are the graphical representations of real-world systems through a collection of nodes joined together in pairs by edges representing a wide variety of relationships. Network analysis offers a powerful analytical tool to study relational data as well as to analyze structural characteristics at individual- and system-levels. Social network analysis (SNA), one of the commonly used methods in social science studies, focuses on analyzing social systems through modelling the social interactions among social actors, such as in friendship networks. SNA enables us to quantify individual and group status based on the structural characteristics of social connections (Wasserman \& Faust, 1994; Newman, 2018). It has been considered an effective approach to analyze social dynamics and interactions among group members in computer-supported collaborative learning (CSCL; Chen \& Poquet, 2022; Saqr et al., 2022; Dado \& Bodemer, 2017; Gašević et al., 2019; Oshima et al., 2012; Reffay \& Chanier, 2003). The applications of SNA in CSCL are centred around analyzing communication-based ties for understanding the information flow among participants as well as identifying active or peripheral actors in online discussion forums and wiki-based collaborative writing (Dado \& Bodemer, 2017; de Laat et al., 2007; Rabbany et al., 2014; Saqr \& López-Pernas, 2022). Kaliisa et al. (2022) highlighted the need for innovative network approaches for social learning analytics.\\

Bipartite network analysis (BNA) models the connections between two types of node sets (i.e., two-mode network), in contrast to traditional social network analysis, which focuses on the relationships among nodes of a single type (Newman, 2018; Wasserman \& Faust, 1994). Bipartite networks model the heterogenous relationships between different types of entities. The heterogenous nature of bipartite networks makes them a powerful and flexible tool for modelling the complex relationships within various social and physical systems (Valejo et al., 2021). In CSCL, a one-mode network can be used to capture the social interactions among group members, with nodes in the network representing students. A bipartite network in CSCL can be used to measure the associations between different types of nodes, such as student–subtask relationships.\\

Currently, there is little usage of BNA in learning analytics and CSCL. Matsuzaw et al. (2011) developed an analytical platform for analyzing word-discourse relationships using bipartite network analysis in order to understand the knowledge-building process among group members in CSCL. Feng et al. (2024) constructed bipartite projections from a heterogenous tripartite network consisting of multimodal process data in small-group collaborative learning to identify the significant behavioural engagement strategies of individual students based on the associations between student communication and spatial movements. This previous research shows that BNA offers an effective tool to model and analyze the structural regularities within heterogenous relationships (e.g., students, artefacts, theory-informed constructs) involved in collaborative learning.\\

Networks are the graphical representations of real-world systems through a collection of nodes joined together in pairs by edges representing a wide variety of relationships. Network analysis offers a powerful analytical tool to study relational data as well as to analyze structural characteristics at individual- and system-levels. Social network analysis (SNA), one of the commonly used methods in social science studies, focuses on analyzing social systems through modelling the social interactions among social actors, such as in friendship networks. SNA enables us to quantify individual and group status based on the structural characteristics of social connections (Wasserman \& Faust, 1994; Newman, 2018). It has been considered an effective approach to analyze social dynamics and interactions among group members in computer-supported collaborative learning (CSCL; Chen \& Poquet, 2022; Saqr et al., 2022; Dado \& Bodemer, 2017; Gašević et al., 2019; Oshima et al., 2012; Reffay \& Chanier, 2003). The applications of SNA in CSCL are centred around analyzing communication-based ties for understanding the information flow among participants as well as identifying active or peripheral actors in online discussion forums and wiki-based collaborative writing (Dado \& Bodemer, 2017; de Laat et al., 2007; Rabbany et al., 2014; Saqr \& López-Pernas, 2022; Chen \& Huang, 2019). Kaliisa et al. (2022) highlighted the need for innovative network approaches for social learning analytics.\\

Bipartite network analysis (BNA) models the connections between two types of node sets (i.e., two-mode network), in contrast to traditional social network analysis, which focuses on the relationships among nodes of a single type (Newman, 2018; Wasserman \& Faust, 1994). Bipartite networks model the heterogenous relationships between different types of entities. The heterogenous nature of bipartite networks makes them a powerful and flexible tool for modelling the complex relationships within various social and physical systems (Valejo et al., 2021). In CSCL, a one-mode network can be used to capture the social interactions among group members, with nodes in the network representing students. A bipartite network in CSCL can be used to measure the associations between different types of nodes, such as student–subtask relationships.\\

Currently, there is little usage of BNA in learning analytics and CSCL. Matsuzaw et al. (2011) developed an analytical platform for analyzing word-discourse relationships using bipartite network analysis in order to understand the knowledge-building process among group members in CSCL. Feng et al. (2024) constructed bipartite projections from a heterogenous tripartite network consisting of multimodal process data in small-group collaborative learning to identify the significant behavioural engagement strategies of individual students based on the associations between student communication and spatial movements. This previous research shows that BNA offers an effective tool to model and analyze the structural regularities within heterogenous relationships (e.g., students, artefacts, theory-informed constructs) involved in collaborative learning.

\subsection{Research Gaps}

\noindent In general, three aspects are not fully explored in the previous studies analyzing students’ emerging roles in CSCL:
\begin{enumerate}
    \item Novel representations for analyzing students’ emerging roles in small group CSCL
    \item Quantitative analysis of the quantity and heterogeneity of individual contributions
    \item Effects of scripted roles on emerging roles
\end{enumerate}

This study introduces a bipartite network approach to gauge students’ emerging roles based on the two dimensions of individual contributions — quantity and heterogeneity — using the interactions between students and the subtasks involved in their small group CSCL processes. Student–subtask engagement data offers a valuable opportunity to acquire an effective understanding of the emerging product-oriented roles in a learning process. Compared to social interaction data, student–subtask engagement data provides direct evidence of cognitive engagement in the process of production. This study utilized a bipartite network approach to capture the student–subtask interactions and further analyze the emerging roles of students based on the quantity and heterogeneity of individual contributions. The importance of the variety or breadth of subtasks engaged in by a student in the learning process is underexamined. Weinberger and Fischer (2006) introduced a multidimensional conceptual framework for understanding argumentative knowledge construction in CSCL, which considers participation from the perspectives of individual participation quantity and group-level heterogeneity. The analysis of individual participation was based on the word count within discourses; the group-level heterogeneity was determined by the standard deviation of individuals’ contributed words within a group. In the current study, we examine both the quantity and heterogeneity at the individual level. In this context, the heterogeneity of individual contribution reflects to what extent a student engages in a variety of subtasks in collaborative learning, rather than the equality of participation among individuals within a group. The heterogeneity of individual contribution examined in this study therefore provides an important lens to understand the breadth of possible cognitive learning opportunities a student encounters during a collaboration (Kawakubo et al., 2022). Lastly, previous studies have mainly focused on investigating the value of scripted roles on students’ cognitive presence and knowledge construction (Gašević et al., 2015; Wise \& Chiu, 2011; De Wever et al., 2010; Rolim et al., 2019), which we have expanded to focus on the impact of scripted roles on emerging roles. This study contributes towards bridging these three research gaps by introducing a bipartite network method for uncovering students’ emerging roles based on an analysis of individual contributions, as well as exploring the effect of assigned leadership roles on their emergent individual contributions.

\section{Methods}
\subsection{Participants and Context}
Twenty-one students from a public high school in Australia were invited to participate in this study. These students attended two elective classes for the same subject with the same course materials and assignments, taught by two collaborating teachers. Students in the two classes were assigned into groups by the teachers and were required to complete two semester-long team project (TP) assignments (TP1 and TP2) throughout an academic year to research, design, and construct a social media platform for informing and training people at a workplace regarding health and safety issues.\\

Teachers assigned a team leader to each group (the scripted role) at the beginning of each project, after which each group was required to self-organize their collaborative process and make their own decisions about the task distribution, coordination, and collaboration. During the first project (TP1), the 21 students were assigned to seven groups of three. After the first project, one student dropped out of the course and the remaining two students were placed into two existing groups for completing the second group project (TP2). Most students remained in the same group during the two projects and played the role of leader and regular group member (e.g., a leader in the TP1 might become a regular team member in TP2). A few students did not have a chance to be a team leader in either project. Students expected to receive a nationwide certification concerning business knowledge training upon the completion of the two projects. A summary of the descriptive information of the two group projects and participants of the seven groups is provided in Table 1 and Table 2.\\

Students completed both projects (TP1 and TP2) using an online collaborative learning platform — a web-based, mobile-ready application platform for active digital learning experiences and event-level data collection (example screenshots of the platform are shown in Figures 1 and 2 below). The platform supports teachers in providing design-based scaffolding for organizing CSCL activities. Authoring on the collaborative learning platform required the two teachers working together to hierarchically decompose the required deliverables of TP1 and TP2 into a series of subtasks, which students were required to collectively complete with their group members. Teachers also categorized the subtasks into three types based on their objectives. Each subtask was assigned a point value by the teachers based on its complexity and difficulty.\\

Instructions to all student teams were delivered in a blended approach with both face-to-face instructions and online interactions. To ensure consistency between the two classes and minimize any influence by teachers on students’ collaborative learning process, several measures were taken. In the study, we did not introduce any interventions to the students’ collaborative learning processes. The individual contributions of students were uniformly and unobtrusively observed via analysis of the digital traces of the team collaborative efforts. The 21 students were in the same grade and took classes of the same subject. The design of the two group projects and the subtasks of each group project were exactly the same. At the beginning of the project, the research team had several meetings with the two teachers to familiarize them with the collaborative learning platform, co-design the collaborative learning projects on the platform, and ensure consistency of teaching methods between the two classes. The two teachers held regular meetings before and during the classes to develop the subtasks of the two group projects together and ensure consistency in the instructions given to students. The two teachers facilitated students’ collaborative learning processes, and students had autonomy to self-organize their two group projects.

\begin{table}[h]
\centering
\caption{Summary of TP1 and TP2}
\label{tab:summary}
\begin{tabular}{|c|c|c|}
\hline
\multicolumn{3}{|c|}{\textbf{General Information}} \\
\hline
 & \textbf{TP1} & \textbf{TP2} \\
\hline
Number of groups & 7 & 6 \\
\hline
Number of students & 21 & 20 \\
\hline
Number of subtasks & 78 & 58 \\
\hline
Subtask point values (with frequency) & 2 (19), 3 (28), 5 (17), 10 (12) & 2 (7), 3 (23), 5 (27), 10 (1) \\
\hline
Types of subtasks & Written (35), Research (26), Design (17) & Written (22), Analysis (31), Logistics (5) \\
\hline
\end{tabular}
\end{table}

\begin{table}[h]
\centering
\caption{Team Composition for TP1 and TP2}
\label{tab:teams}
\begin{tabular}{|c|c|c|c|}
\hline
\multirow{2}{*}{\textbf{Class}} & \multirow{2}{*}{\textbf{Team}} & \multicolumn{2}{c|}{\textbf{Students}} \\
\cline{3-4}
 & & \textbf{TP1} & \textbf{TP2} \\
\hline
\multirow{2}{*}{Class\_1 (Teacher\_1)} & Team\_1 & S1,3,7 & S1,2,3,7 \\
\cline{2-4}
 & Team\_2 & S13,15,5 & S13,14,15,5 \\
\cline{2-4}
 & Team\_3 & S16,20,21 & S16,20,21 \\
\cline{2-4}
 & Team\_4 & S14,2,6 & S6 dropout \\
\hline
\multirow{2}{*}{Class\_2 (Teacher\_2)} & Team\_5 & S12,17,4 & S12,17,4 \\
\cline{2-4}
 & Team\_6 & S10,11,19 & S10,11,19 \\
\cline{2-4}
 & Team\_7 & S18,8,9 & S18,8,9 \\
\hline
\end{tabular}
\end{table}

\begin{figure*}
    \centering
    \includegraphics[width=\textwidth]{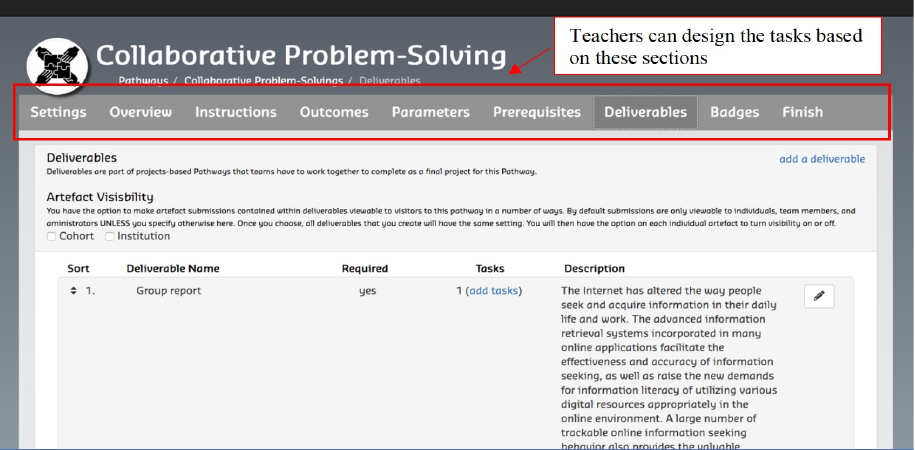}
    \caption{A screenshot of the teacher interface on the collaborative learning platform. Teachers define the deliverables of TP1 and TP2 as well as the subtasks before starting a group project.
    }
    \label{fig:diagram}
\end{figure*}

\begin{figure*}
    \centering
    \includegraphics[width=\textwidth]{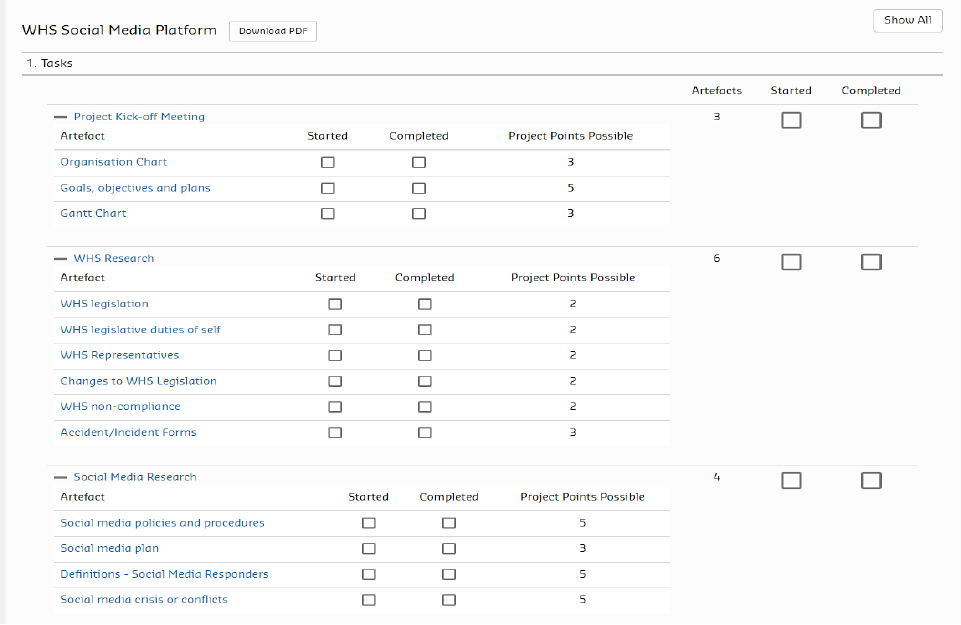}
    \caption{A screenshot of a subset of subtasks for the first project (TP1). Students are required to self-organize the collaborative learning process to complete all subtasks of each project on the collaborative learning platform.
    }
    \label{fig:diagram}
\end{figure*}

\subsection{Data Analysis}
\subsubsection{The Student–Subtask Bipartite Network}
Bipartite network analysis is used to analyze the relationships between two different sets of nodes in a network (Newman, 2018). Within the context of CSCL, a bipartite network in this study is constructed to model the interactions between students and subtasks in each group in both projects, based on the process data retrieved from a collaborative learning platform. An edge in the bipartite network between the two sets of nodes represents the association of a student and a subtask that they worked on (Figure 3). Two adjusted network measures were developed in this study for analyzing the quantity and heterogeneity of individual student contributions based on the constructed student–subtask bipartite networks.\\

\begin{figure*}
    \centering
    \includegraphics[width=0.5\textwidth]{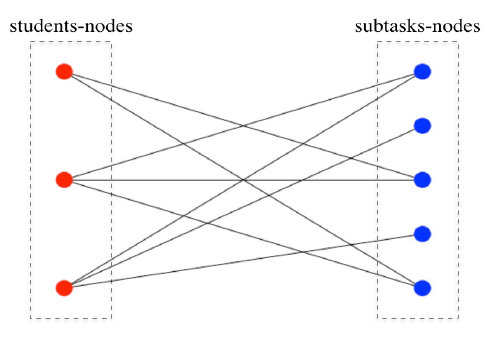}
    \caption{Illustration of the student–subtask bipartite network.
    }
    \label{fig:diagram}
\end{figure*}

\subsubsection{Analysis of the Quantity and Heterogeneity of Individual Contributions (RQ1)}

\hspace{-15pt}
\emph{2.1 Quantity of individual contribution}

First, an adjusted degree measure was developed to assess the quantity of individual contributions, taking the grade weight of each subtask into consideration (Eq. 1). As multiple students can work on the same subtask together in collaborative learning, the quantity of individual contribution in this study assesses the weighted number of subtasks a student participates in during the learning process. Degree centrality in SNA normally gauges the number of edges of a node in a network (Scott, 1988). In a bipartite network, without transforming the bipartite network into two one-mode networks, degree centrality is the numerical measure of the number of connections that a node has with the nodes in the other set. This measure answers questions such as “How many subtasks has a student worked on?” Within the context of CSCL, the (normalized) degree centrality of the nodes in the student set measures the fraction of subtasks a student has engaged with in the bipartite network. More concretely, the degree centrality $d_i$ of a node $i$ in the student set $V_s$ in this bipartite network is the normalized number of subtasks $j$ in the subtask set $V_a$ worked on by the student $i$ in the group collaboration
\begin{align}
d_i = \frac{\sum_{j\in V_a}e_{ij}}{\abs{V_a}},    
\end{align}
\noindent where $e_{ij}$ is the number of edges (0 or 1) between a node $i$ in the student set $V_s$  to a node $j$ in the subtask set $V_a$.\\

It is also important to take the complexity and difficulty of subtasks into consideration while computing the quantity of individual contributions, allowing us to analyze to what extent each group member has worked on more critical subtasks. The point values assigned by teachers for subtasks were used to generate the weighted degree centrality $dw_i$ of a node $i$ in the student set and represent the weighted average number of subtasks $j$ that a student $i$ contributed to, where the weight $w_j$ is the point value assigned to subtask $j$ by the teacher (Eq. 2).
\begin{align}
 dw_i = \frac{\sum_{j\in V}e_{ij}w_j}{\sum_{j\in V}w_j}.   
\end{align}

Based on Eq.~2, a weighted degree value closer to 1 indicates that the individual student contributes to more subtasks and/or subtasks with higher point values in the project, and a value closer to 0 indicates that the individual student contributes to fewer subtasks and/or subtasks with lower point values. In this study, we used the weighted degree centrality to measure the quantity of individual contributions in CSCL, considering both the number of subtasks as well as the point values of the subtasks contributed to (but not necessarily completed) by an individual student in the learning process.\\

\hspace{-15pt}
\emph{2.2 Heterogeneity of individual contribution}

One of the goals of collaborative learning is to help individuals to learn better in a collective setting. Assessing the quantity of individual contribution can contribute to understanding how much an individual student contributes to a group collaboration. In addition to the quantity, a measure of the heterogeneity of individual contribution aims to assess to what extent students participated in diverse subtasks in the collaborative learning process. In this study, we adapted an entropy-based measure used in Feng and Kirkley (2020), based on information theory (Shannon, 2001), to assess the heterogeneity of subtasks $H_i$ that an individual student $i$ is involved in within a group collaboration, based on the subtask categories in a project (Eq. 3).
\begin{align}
   H_i &= -\frac{1}{Q(n_i)}\sum_{t=1}^{T}\frac{n_{it}}{n_i}\log \left(\frac{n_{it}}{n_i}\right),\\
   Q(n_i) &= \max_{\{n_{it}\}_{t=1}^{T}}\left\{-\sum_{t=1}^{T}\frac{n_{it}}{n_i}\log \left(\frac{n_{it}}{n_i}\right)\right\}.
\end{align}

In Eq. 3, $n_{it}$ is the number of student $i$’s subtasks under the task type $t$, $T$ is the total number of subtask types in the group project ($T=3$ in this case), and $n_i$ is the number of tasks student $i$ participated in. The base of the logarithm here does not change the result. The factor $Q(n_i)$ is a normalization constant that enforces the measure $H_i$ to lie in the range $[0,1]$, and is computed by distributing the $n_i$ tasks as evenly as possible across the $T$ task types prior to plugging into the summation, using the total number of subtasks of each task type. For example, in our case $n_i = 21$ subtasks, so we let $n_{it}=7$ for all $t$, giving $Q(n_i)=\log 3$. However, if the number of total subtasks is much larger than the number of subtasks of a particular type, e.g. if $n_i=57$, then student $i$ cannot possibly contribute equally to all task types $t$ as there were only $17$ ``Design'' tasks while $57/3 = 19$ is required for equal numbers of each type. In this case, the maximum value of $Q(n_i)$ occurs when the $n_{it}$ values are $20$, $20$, and $17$ respectively for ``Writing,'' ``Research,'' and ``Design,'' and this distribution results from pedagogical decisions made by the teamwork designer.\\

To identify students’ emerging roles based on the two conceptual dimensions (Eq. 1 and Eq. 2), a predefined value of $0.5$ was assumed for both measures (considering their range of $[0,1]$) as a threshold for classifying students’ emerging roles into different categories. We acknowledge that data-driven clustering approaches have been commonly adopted to identify the emerging roles of students in previous studies (Heinimäki et al., 2021; Li et al., 2024; Mi et al., 2023). However, the constraints of data-driven clustering analysis lie in the unstable number of clusters depending on the dataset structure and the subjective interpretation of the informed clusters. The inconsistent number and composition of clusters across different datasets make it challenging to compare students’ emerging roles across multiple collaborative learning projects. Additionally, the subjective interpretation for labelling each cluster could lack a clear justification and theoretical basis. Therefore, in the current study, we employed a predefined threshold that allows us to have a stable classification of students’ emerging roles for analyzing role changes across the two consecutive collaborative learning projects as well as a principle-guided interpretation of the resulting clusters. However, it is important to note the trade-off of using a pre-defined threshold — it may mask the differences among the individuals with borderline values and those exhibiting extremely low or high values in either dimension. The threshold can be adjusted based on new theoretical considerations or pedagogical demands in teaching practices while using the proposed method in other studies, illustrating the flexibility of the proposed method. Future studies can also adopt the measures proposed in this study and use data-driven clustering analysis methods to identify emerging roles depending on the analytical needs.\\

\subsection{Analysis of the Effects of Assigned Leadership Roles on Emerging Roles (RQ2)}
A two-sample Mann-Whitney U-test was used to analyze the difference in the distribution of individual contributions between leaders and group members in the two projects. The analysis aimed to reveal whether there was an observable and meaningful difference in the emerging roles, informed by the new measures of individual contributions, when comparing students with and without an assigned leadership role. In addition, Barnard’s test was used to further examine the associations of changes in assigned scripted roles with the changes in students’ emerging roles. Barnard’s test is considered a more robust and powerful alternative to Fisher’s exact test for assessing significance in contingency tables with small sample sizes (Barnard, 1947; Andrés et al., 2004).

\subsection{Analysis of Qualitative Interview Data for Validating the Quantitative Results (RQ3)}

A semi-structured interview of individual students was conducted after the completion of the first project to help identify qualitative evidence for supporting the analysis results. Eleven students from four groups participated in the interview voluntarily. Individual interviews took 20 minutes for each participant and were audio-recorded. Students were asked about how they see their own performance in the collaborative learning process as well as how much they learned through the project.\\

The transcript of the interview data was stored in a password-protected laptop of the primary investigator. The transcribed interview data was read several times to highlight the meaningful content under each question in all responses. Second, the research team merged the open codes for each question into several categories. The last step was to merge the codes under each category into themes that reflect the factors associated with individual contributions. This process was completed independently by two researchers and results were compared. Student perceptions of their emerging roles based on the interview data were compared with the bipartite network analysis results. The additional insights related to the causes and effects of different emerging roles in the learning processes were also reported. This process of weaving qualitative and quantitative phases (Creswell et al., 2003; Miles \& Huberman, 1994), and validating with independent observers was further extended by sharing the results with the two teachers for confirmation of validity as well as for gaining additional insights into how to help students become effective individual learners in collaborative learning.

\section{Results}
\subsection{Emerging Roles Based on the Quantity and Heterogeneity of Individual Contributions (RQ1)}

Network visualizations of the student–subtask bipartite networks provide an effective way to gain an intuitive understanding of students’ emerging roles within a team (see an example of Team 3 in Figure 4 below). In Team 3, student S16 was assigned as the team leader in the first project and then student S20 was assigned as the team leader in the second. Figure 4 reveals that while S16 was the assigned team leader, they worked on a higher portion of the subtasks in the first project than other members. Meanwhile, in the second project, after stepping down from the leadership role, student S16 worked on relatively few subtasks and S20, the newly assigned team leader, worked on most of the subtasks. This network visualization can help teachers and students get a quick understanding of the overall performance of each student in a CSCL processes. However, it would be challenging to process and compare many students and groups using solely qualitative inspection of network visualizations. This study validates that the two new bipartite network measures (Eq. 1 and Eq. 2) can correctly analyze the quantity and heterogeneity (variety/breadth) of individual student contributions in this small group CSCL context.

\begin{figure*}
    \centering
    \includegraphics[width=\textwidth]{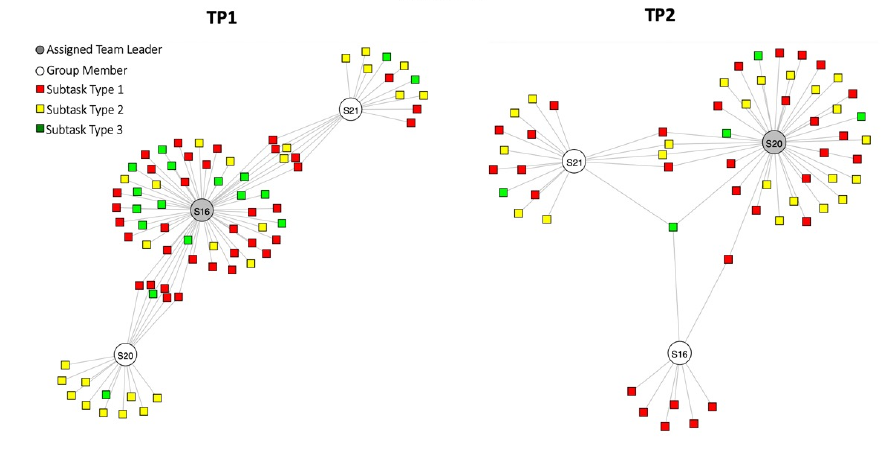}
    \caption{Bipartite network visualizations of student–subtask interactions for Team 3 in TP1 and TP2.
    }
    \label{fig:diagram}
\end{figure*}

Using the method outlined, the study identified three types of emerging roles: 1) comprehensive contributors (high quantity and high heterogeneity), 2) versatile participants (low quantity and high heterogeneity), and 3) free riders (low quantity and low heterogeneity; see Figure 5). Comprehensive contributors had a high heterogeneity and high quantity of individual contribution, exemplifying the most desirable situation for a student’s contributions in a collaborative learning activity. Versatile participants had a high level of heterogeneity but a low quantity of individual contributions, contributing sparsely to a variety of subtasks. Free riders fell into the category of low heterogeneity and low quantity. The identified free riders should raise the attention of teachers, as these students are expected to have the least individual learning gains from the group collaboration (Cohen \& Lotan, 2014).\\

As an example, consider Team 3 shown in Figures 4 and 5. Here, student S16 is identified as a comprehensive contributor in the first project (TP1), while the other two group members, S20 and S21, are identified as versatile participants who did not contribute prolifically but worked on all three types of subtasks with relatively balanced efforts. In the second project (TP2), S20, who was assigned the role of leader, is identified as a comprehensive contributor, while S21 is identified as versatile participant and S16 is identified as a free rider. Compared to S20 and S21, the participation of S16 mainly focused on a single type of subtask, which could preclude the student from co-constructing the knowledge associated with the other types of subtasks. This further highlights the importance of using the heterogeneity measure in understanding individual student contributions and knowledge gains in small group CSCL.\\

From Figure 5, we can observe that all students with an assigned leadership role from the seven groups in the first project (red stars) were comprehensive contributors. Student S18 from Team 7 was the only group member (not an assigned leader) who was a comprehensive contributor in TP1. Meanwhile, student S9 from Team 7 was the only free rider in the first project. Most students with no assigned leadership roles in the first project were identified as versatile participants. In the second project (TP2), we observed a decline of the quantity of individual contribution among all students and an increase in the number of free riders compared to TP1. In this period, among the six students with assigned group leadership roles (blue stars), three were identified as comprehensive contributors, while the other three were identified as versatile participants. Most students with no leadership role fell under the category of versatile participants; an increased number of free riders were identified in the second project. The observed associations between assigned leadership roles and emerging roles in Figure 5 led us to examine the relationships rigorously in Section 5.2. The analysis of student interview responses in section 5.3 also provides detailed insights about the observed patterns and performance.

\begin{figure*}
    \centering
    \includegraphics[width=0.6\textwidth]{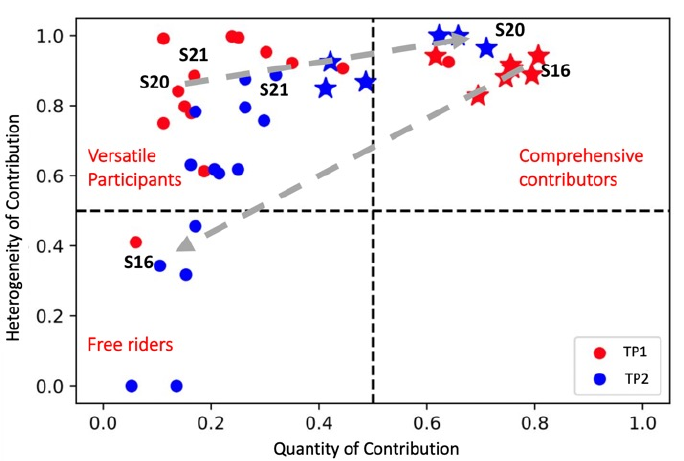}
    \caption{The analysis results of the emerging roles based on the quantity and heterogeneity of individual contributions among the same cohort of students over two projects. The dashed grey arrows indicate the transition in emerging roles from the first team project (TP1) to the second team project (TP2) for students S16 and S20 from Team 3 (see Figure 4).
    }
    \label{fig:diagram}
\end{figure*}

\subsection{The Effects of Assigned Leadership Roles on Individual Contributions (RQ2}

We first analyzed the effects of assigned leadership roles on the quantity and heterogeneity of individual student contributions respectively using a two-sample Mann-Whitney U-test for small sample sizes. The quantity of individual contributions in the first project was significantly greater for leaders ($\text{Mdn} = 0.755$, $N=7$) than for non-leaders ($\text{Mdn} = 0.178$, $N=14$), $U = 1$, $p = .0001$, $r = 0.781$. The same result was found in the second project. The quantity of individual contributions in the second project was significantly greater for leaders ($\text{Mdn} = 0.555$, $N=6$) than for non-leaders ($\text{Mdn} = 0.189$, $N=14$), $U = 0$, $p = .0003$, $r =0.775$. The results suggest that assigned group leadership roles had a positive influence on the quantity of individual contributions.\\

A two-sample Mann-Whitney U-test was also performed to analyze the difference in the distribution of the heterogeneity of individual contribution between leaders and group members in the two projects. The heterogeneity of individual contributions in the first project was not significantly greater for leaders ($\text{Mdn} = 0.906$, $N=7$) than for non-leaders ($\text{Mdn} = 0.896$, $N=14$), $U = 45$, $p = .397$, $r = 0.065$. However, the heterogeneity of individual contributions in the second project was significantly greater for leaders ($\text{Mdn} = 0.944$, $N=6$) than for non-leaders ($\text{Mdn} = 0.618$, $N=14$), $U = 4$, $p = 0.001$, $r = 0.701$. Even though the distribution of types of subtasks in the three types (written, analysis, and logistics) in the second project was more uneven than in the first project (see Table 1), we accounted for this factor in the proposed heterogeneity measure (Eq. 3) through normalization.\\

The Mann-Whitney U-test analysis results, together with observations based on Figure 5, indicated that there were significant differences in the quantity and heterogeneity of individual contributions between students with and without assigned leadership roles. To further examine the effects of assigned leadership role on students’ emerging roles, we analyzed the associations between the changes in students’ emerging roles and the changes in assigned leadership roles between TP1 and TP2. Figure 6 presents a $2 \times 2$ contingency table showing the number of students with changes in the assigned leadership roles and the number of students with changes in the identified emerging roles after the first project. As one student dropped the class after the first group project, there were a total of 20 students analyzed in Figure 6. It shows that most students changed their emerging roles in TP2 relative to TP1 (13 out of 20), while nine out of 20 students in total either stepped down or assumed new leadership roles in the second project. Among the nine students who changed their leadership roles, eight students’ emerging roles changed. Barnard’s test was applied to analyze the association in the changes. We found that there was a significant association in the changes between assigned leadership roles and changes in the identified emerging roles ($T = 2.026$,$N = 20$,$p = 0.05$). This provides further support for the association between assigned leadership roles and emerging roles as well as the importance of externally facilitated regulation scaffolding in CSCL processes (Gašević et al., 2015).

\begin{figure*}
    \centering
    \includegraphics[width=0.6\textwidth]{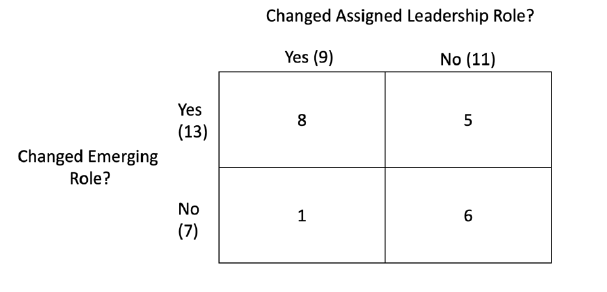}
    \caption{Changes of Assigned Leadership Roles and Identified Emerging Roles from the First to the Second Project).
    }
    \label{fig:diagram}
\end{figure*}

\subsection{Student Perceptions of Their Emerging Roles (RQ3)}
Eleven students participated in the individual interviews voluntarily after the first project, including S10, S19, S11 (Team 6), S8, S18, S9 (Team 7), S20, S21, S16 (Team 3), S2 and S14 (Team 4). Among these eleven students, S10, S14, S8, and S16 were identified as the comprehensive contributors based on the bipartite network analysis, and these students were also the assigned team leaders in the first project. Student S18 was the only regular group member identified as a comprehensive contributor in the first project based on our quantitative analysis results. The emerging role of S9 in the first project was a free rider, and S2, S19, S11, S20, and S21 were identified as versatile participants.\\

The final percentage agreement in the coding results of the interview data by the two researchers based on the formula by Miles and Huberman (1994) was 0.92. We first examined student perceptions of their contributions in the first project based on their responses to the following questions: What is your role in the first project? And how do you see your performance? The students who were leaders and identified as the comprehensive contributors had a clear awareness of their assigned leadership role (see, for example, Comment 1 in the supplementary document, hereafter referred to as SD). The comprehensive contributors indicated that they put a large amount of time and effort into the first project (see Comment 2 SD). S18, the only comprehensive contributor without an assigned leadership role, shared that they valued the national certificate to be awarded for completing the course. This student indicated a very positive attitude towards group projects and expressed their devotion to the project (see Comments 2 and 3 SD).\\

The students who were identified as versatile participants all indicated a level of uncertainty about their roles in the first project in their responses (see Comment 7 SD). A direct quote from student S20 is ``so I did some of the research, I did some of the summaries. So, we basically did like, a bit of everything.'' This response provided an example of a versatile participant who contributed to a bit of everything, corroborating the analytical results as well as the conceptual definition of versatile participants. In addition, the versatile participants indicated uncertain feelings and a lack of confidence about their group project at the beginning due to such reasons as confusion about the project, unfamiliarity with their group members, the pace of early progress, and a high tendency to get distracted (see Comments 9 and 10 SD). Different from the versatile participants, some comprehensive contributors with an assigned leadership role indicated a high level of confidence about the group project at the beginning (see Comment 6 SD). This result is consistent with the previous studies finding that student confidence in their own ability plays a role in affecting performance in CSCL (Wilson \& Narayan, 2016). Teachers should provide support and interventions to strengthen student confidence and clarify uncertainties with versatile participants in particular, so they can be better supported and become comprehensive contributors to the project.\\

Through the interview responses, we also gained insights about individual student gains through the first project. There are three types of individual gains mentioned by students, including 1) teamwork experience; 2) subject relevant knowledge; and 3) digital technology skills, such as the use of Microsoft Office or the collaborative learning platform. Most students, regardless of their emerging roles, shared that teamwork experience was the best part of the project, followed by the sharing of subject relevant knowledge (see an example in Comment 5 SD). The students identified as comprehensive contributors in the first project indicated all three types of individual gains, while the versatile contributors only mentioned one or two gains. This finding provides insights about the relationships between individual contributions and individual gains in CSCL (Chen et al., 2018).
Two identified comprehensive contributors (S10 and S14) mentioned that the best part of the project was that the online collaborative learning platform makes group communication easy and provides hints to support the learning process (see Comments 5 and 6 SD). This suggests that students appreciated the online learning platform used to support their collaborative learning process and find it useful for helping them to communicate with teachers as well as receive timely guidance. This provides an insight that digital technologies not only facilitate the development and implementation of collaborative learning activities but also provide an opportunity for students to be exposed to digital learning systems and help students to develop their digital literacy (Ludvigsen et al., 2018).\\

In general, we found that the qualitative analysis results based on student interview responses corroborate the bipartite network analysis results. They also provide additional in-depth insights to understand the characteristics of each emerging role in the collaboration processes.

\section{Discussion}
\subsection{Bipartite Network Analysis for Collaboration Analytics}
Being able to quantitatively measure individual contribution using advanced learning analytics methods and fine-grained data in the learning process of CSCL is one of the goals of collaboration analytics (Schneider et al., 2021). The individual aspect tends to be neglected in CSCL, as it is challenging to disentangle individual performance from group-based deliverables (van Aalst, 2013). This study demonstrates that bipartite network analysis provides an effective method for analyzing students’ emerging roles based on student–subtask interactions. Different from one-mode networks that model the relationship among the same type of nodes, bipartite network analysis can capture complex relationships between two different types of nodes, in this case students and subtasks.\\

The bipartite network analysis described and validated in this study can provide individual-level information to teachers and students during the learning process, which is recommended to be integrated into future learning platforms for supporting teachers to make timely and well-informed decisions as well as for helping students to reflect on their performance and make adjustments. In addition, this study also demonstrated that the visualization of bipartite networks (Figure 4) provides an effective way to help teachers and students gain a quick understanding of the relationships between students and subtasks in the learning process. A network-based visualization can also be used as an intervention tool to guide students to enhance their metacognitive awareness of their contributions to strengthen their regulated behaviour in CSCL processes (Banihashem et al., 2022; Gašević et al., 2015).\\

It is worth noting that in this study we did not transform the constructed bipartite networks into one-mode networks, a technique commonly used in social network analysis to enable the application of widely used network measures. Our rationale for not transforming the bipartite network is that we aimed to model student–subtask interactions in small-group collaborations, rather than analyze the similarities among students or subtasks alone. In the context of the student–subtask bipartite networks of this study, the one-mode networks might be useful to indicate the co-work relationships among students on the same subtasks, or the co-ownership of subtasks among the same students, which are not directly relevant for addressing the research questions of this study. For analyzing similarities among students or tasks in large-group contexts, future studies using bipartite network analysis may consider projecting the network onto one-mode networks for further analysis.

\subsection{A New Framework for Understanding Emerging Roles in Small Group CSCL}
Strijbos and De Laat (2010) proposed four types of emerging roles in small group CSCL, including over-rider, ghost, free rider, and captain, along the dimension of orientation and efforts. Strijbos and De Laat (2010) used individual stances towards collaborative learning to gauge students’ emerging roles in the learning processes. The dimension of orientation refers to the value of individual student goals or group goals, and the dimension of effort refers to product-oriented student investment as well as process-oriented investment in the collaborative assignments. Different from Strijbos and De Laat’s approach, we use quantified product-oriented contributions to gauge students’ emerging roles in small group CSCL. We identified and validated three types of emerging roles (RQ1, RQ3) in this study: comprehensive contributor, versatile participant, and free rider. \\

Comprehensive contributor is considered a desirable emerging role in CSCL as it characterizes the students that make a significant effort as well as have thorough learning experiences. Comprehensive contributors shared the traits of ``Captain'' proposed by Strijbos and De Laat (2010) in that they invest much effort in their group task. The high heterogeneity and quantity of contributions reflect their breath and substance of cognitive engagement in the process. In contrast to the comprehensive contributors, versatile participants have a broad but not substantial contribution to their group projects. The last type of emerging role identified in this study is the free rider, who expended minimum effort in the process, as evidenced by both low quantity and heterogeneity of individual contributions. A collaboration without free riders is considered a desirable situation, as free riding is an issue affecting not only group cohesiveness and effectiveness of collaborative learning (Le et al., 2018) but also individual academic gains.
Through the analysis of RQ3, this study found that the students who contributed to the group project significantly in both quantity and heterogeneity reflected that they not only gained teamwork experience and subject-relevant knowledge, but also digital knowledge relevant to the collaborative learning platform. Students with a lower level and breadth of individual contributions only mentioned one or two types of gains through the group project. This finding supports the claim that the occurrence of effective learning at the individual level in CSCL requires active engagement in contributing to group tasks in the learning process (Nokes-Malach et al., 2015).\\

Based on the two dimensions of quantity and heterogeneity of product-oriented contributions (Eq. 1 and Eq. 2) and our analysis results, there are four zones classifying students’ emerging roles. Apart from comprehensive contributors, versatile participants, and free riders, there is a type of emerging role that conceptually exists but was not found in this empirical study, which is the role corresponding to a high quantity and low heterogeneity of individual contribution. Students who work on a significant number of subtasks of similar types would be classified as having high quantity and low heterogeneity and be labelled as “specialized contributors.” This might indicate that the students have a strong preference for, or skills related to, particular types of subtasks in a group project, but have limited their opportunities for other cognitive engagement involved in the project. Future studies using the bipartite network are needed to establish whether the measures provide a comprehensive framework for classification (Figure 7). The conceptualization of the four emerging roles under our framework are presented below:

\begin{itemize}

\item Comprehensive contributors: the individuals who contribute frequently (high quantity) across diverse types of subtasks (high heterogeneity), demonstrating broad and significant involvement in a collaborative learning process.  
\item Specialized contributors: the individuals who contribute frequently (high quantity) but focus narrowly on specific types of subtasks (low heterogeneity), often excelling in a particular area in a group project without diversifying their contributions. 
\item Versatile participants: the individuals who contribute sporadically (low quantity) but spare their efforts across various types of subtasks (high heterogeneity), showcasing broad participation without deep engagement. 
\item Free riders: the individuals who contribute minimally (low quantity) in limited types of subtasks (low heterogeneity), offering little value to a collaborative learning process.

\end{itemize}

\begin{figure*}
    \centering
    \includegraphics[width=0.6\textwidth]{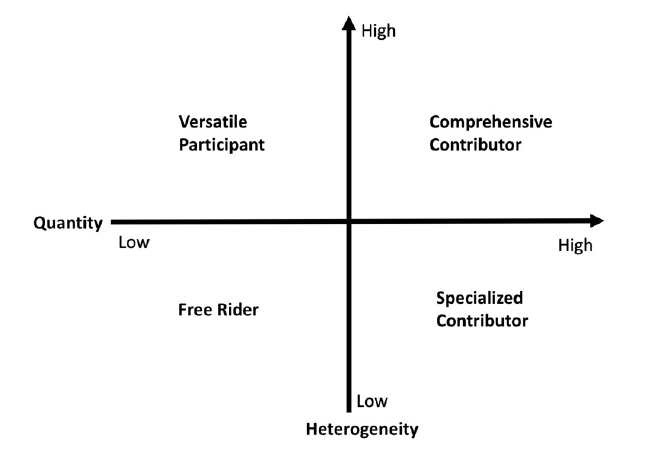}
    \caption{A conceptual framework of students’ emerging roles based on the bipartite network measures of quantity and heterogeneity of individual contributions in CSCL.
    }
    \label{fig:diagram}
\end{figure*}

The four emerging roles provide a theoretical framework for identifying students’ emerging roles in CSCL, as well as for examining the dynamics of emerging roles over time within and across projects. It is worthy to note that future studies that utilize a data-driven approach with the two network measures could uncover new roles based on the number of clusters identified in the dataset. In this case, the conceptual framework with the quantity and heterogeneity dimensions still offers an important theoretical basis for labelling such data-driven clusters. We discussed the pros and cons of data-driven and predefined thresholds for identifying emerging roles in the Data Analysis section. A data-driven approach could result in an inconsistent number of clusters across different datasets, which presents challenges for comparing students’ emerging roles across multiple collaborative learning projects. Future studies can apply our bipartite network measures and conceptual framework with either predefined thresholds or a data-driven clustering method based on their analytical needs.

\subsection{Scripted Roles and Emerging Roles in Small Group CSCL}

Based on the bipartite network analysis results for identifying students’ emerging roles, we also examined the effects of scripted roles — in particular, the assigned leadership roles — on students’ emerging roles in small group CSCL (RQ2). We found a significant association in the changes between assigned leadership roles and changes in the identified emerging roles between the two projects. The impact of the formal leadership role is validated by the trajectories of students who either were non-leader members that became leaders in the second project or leaders in the first project who became regular team members in the second project (RQ2). In all cases, the assigned leadership role was associated with increased performance in both quantity and heterogeneity. Some students who never had the chance to lead performed at the lower end on all measures. This indicates that the leadership role may be an effective factor positively affecting individual student contributions in a collaborative learning process. This result echoes the findings of previous studies that suggested role assignment influenced the collaborative learning process (De Wever \& Strijbos, 2021;  Fransen et al., 2013; Gašević et al., 2015; Wise \& Chiu, 2011; Lim \& Liu, 2006).\\

It is worth noting that there is no competition of individual contribution among group members in collaborative learning. Given the positive association between individual contribution and individual gains in CSCL, the ideal situation of collaborative learning is that all group members contribute to the group project significantly in both quantity and diversity. More collaboration among group members should be encouraged and rewarded, instead of allowing passivity through work specialization, where students are completing subtasks individually and assembling individual work at the group level. To encourage all group members to actively contribute to the group project requires social support from group members, directed efforts from the emerging or assigned team leaders, support and guidance from teachers, and intrinsic motivation from the students (Deci \& Ryan, 2008; Järvelä \& Järvenoja, 2011).

\section{Conclusion}
This study analyzed students’ emerging roles based on bipartite network measures of the quantity and heterogeneity of individual contributions in small group online collaborative learning, examined the effects of assigned leadership roles on students’ emerging roles, and provided qualitative evidence for validating the quantitative results. Based on the network measures of individual contributions, we identified three types of emerging roles: comprehensive contributors, versatile participants, and free riders. Assigned leadership roles had a positive influence on students’ emerging roles. The qualitative analysis results corroborate the bipartite network analysis results and also provide additional insight about individual student contributions and individual gains in CSCL.\\

The contributions of this study are threefold. First, this study contributes new learning analytic methods for analyzing emerging roles and individual contribution in small group online collaborative learning. The new method utilized and partially validated in this study can enrich the measures for collaboration analytics to better leverage the fine-grained data in small group CSCL. Future studies are recommended to further apply and test the proposed measures to analyze student–subtask interactions in CSCL. Second, the study’s results contribute to a new conceptual framework for understanding students’ emerging roles, as well as theoretical insights into understanding the effects of scripted roles on emerging roles in CSCL. Third, the findings of this study provide insights for instructional design and pedagogical methods for teachers to better support collaborative learning activities in practice.\\

The effect of the leadership role on individual student contributions provides an insight that teachers and instructional designers are recommended to consider when making a use of the leadership role for encouraging students to participate in collaborative learning activities. It is important to note that the goal of assessing individual contribution in CSCL is gaining a better understanding of the occurrence of learning at the individual level, rather than encouraging students to complete subtasks individually for a higher individual contribution in group collaboration. Teachers play an important role in providing support and guidance to steer group leaders and members to develop a shared group identity. They can employ novel pedagogical methods such as sharing the learning analytic results with students to let them reflect on their behaviour and develop shared responsibilities within a group.\\

Several limitations of the study can help to set the future research agenda for understanding individual student contributions in collaborative learning processes via bipartite network measures. First, in this study, eleven students joined the interviews voluntarily. It is reasonable to consider the possibility of self-selection bias in this case. However, the eleven students interviewed possessed a variety of scripted and emerging roles, which enabled us to capture a range of viewpoints for addressing RQ3. Second, we only identified three types of emerging roles in the limited empirical data. With a larger dataset, the characteristics of all four types of emerging roles can be further explored. Third, this study examined students’ emerging roles in a post-hoc collection of collaboration data. Future studies are recommended to use the method to examine the dynamic trajectory of evolvement of students’ emerging roles over time in small group CSCL contexts, and to introduce interventions at timed intervals (e.g., pre-data collection, during collaboration, post-collaboration) to examine the effects on the network representations and measures. Lastly, the design of subtasks is critical for the accurate reflection of students’ emerging roles based on this study’s student–subtask engagement data. For example, if a student is not assigned to the role of team leader, how can quantity and heterogeneity of task engagement be encouraged or enhanced? As is the case for all learning analytic studies, the analytical results are influenced by the design of the learning activities. Therefore, we encourage future studies in which a multi-modal approach is used to investigate the multifaceted nature of individual contributions in collaborative learning including the intangible and socio-oriented contributions as well as the task-oriented quantity and heterogeneity of individual contribution tackled in this study. Such a series of studies might also push the boundary of bi-, tri- and multi-partite network representations and force rethinking of the role of nodes and relationships.

\section{Funding}
This project is supported by the Research Grants Council (Hong Kong) under Early Career Scheme \#27605223 (SF).

% \nocite{*}
% \bibliographystyle{apalike}
% \bibliography{refs}

%%%%%%%%%%%%%%%%%%%%%%%%%%%%%%%%%%%%%%%%%%%%%
%%%%%%%%%%%%%%%%%%%%%%%%%%%%%%%%%%%%%%%%%%%%%
\clearpage
\appendix
\onecolumngrid

\begin{figure}[htbp]
  \begin{adjustbox}{width=1.25\textwidth,center}
    \includegraphics{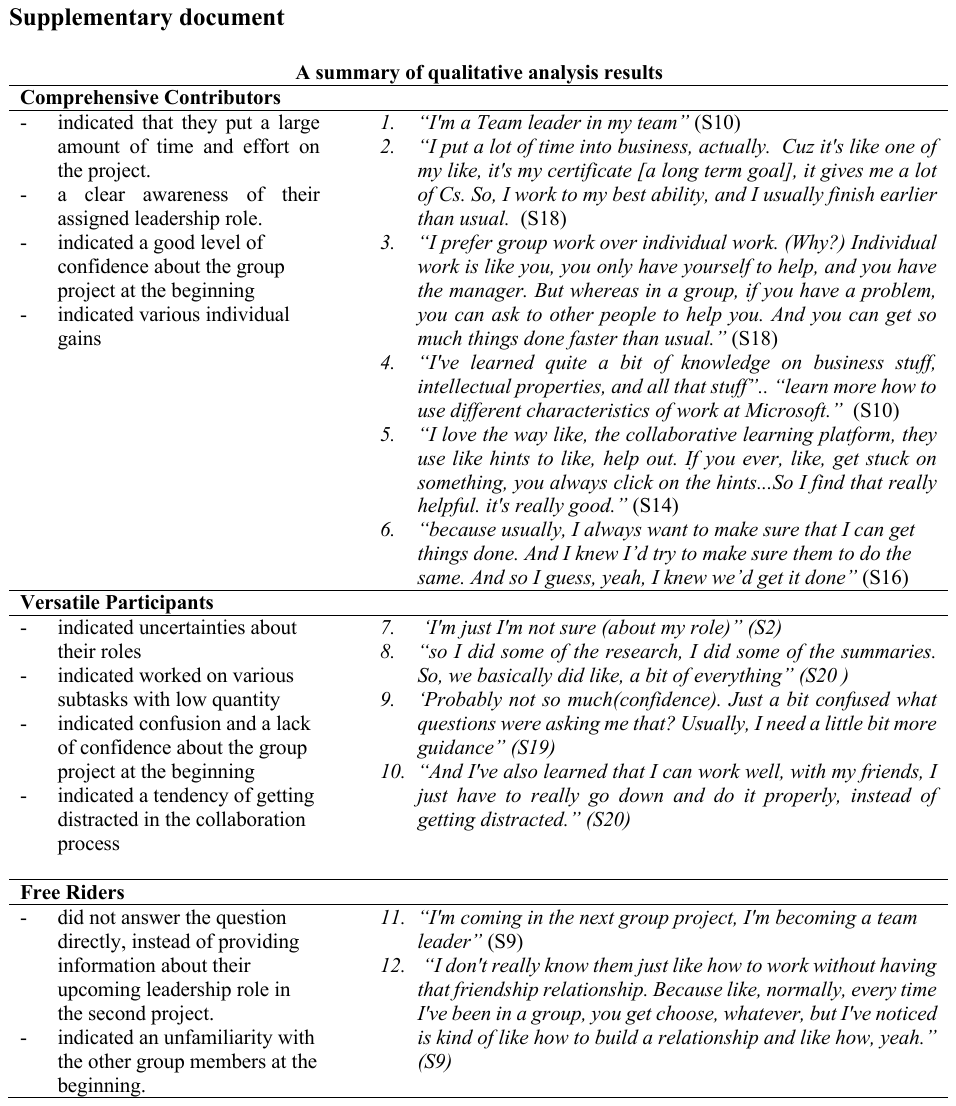}
  \end{adjustbox}
  \label{fig:wide}
\end{figure}

\end{document}